# Analysis of hydrogen permeation tests considering two different modelling approaches for grain boundary trapping in iron

A. Díaz[1*], I.I. Cuesta[1], E. Martinez-Pañeda[2,3], J.M. Alegre[1]

[1]Structural Integrity Group, Universidad de Burgos, Avda. Cantabria s/n, 09006 Burgos. SPAIN

[2]University of Cambridge, Department of Engineering, Trumpington Street, Cambridge CB2 1PZ, UNITED KINGDOM

[3]Department of Civil and Environmental Engineering, Imperial College London, LondonSW7 2AZ, UNITED KINGDOM

[*] Contact e-mail: adportugal@ubu.es

## ABSTRACT

The electrochemical permeation test is one of the most used methods for characterising hydrogen diffusion in metals. The flux of hydrogen atoms registered in the oxidation cell might be fitted to obtain apparent diffusivities. The magnitude of this coefficient has a decisive influence on the kinetics of fracture or fatigue phenomena assisted by hydrogen and depends largely on hydrogen retention in microstructural traps. In order to improve the numerical fitting of diffusion coefficients, a permeation test has been reproduced using FEM simulations considering two approaches: a continuum 1D model in which the trap density, binding energy and the input lattice concentrations are critical variables and a polycrystalline model where trapping at grain boundaries is simulated explicitly including a segregation factor and a diffusion coefficient different from that of the interior of the grain. Results show that the continuum model captures trapping delay, but it should be modified to model the trapping influence on the steady state flux. Permeation behaviour might be classified according to different regimes depending on deviation from Fickian diffusion. Polycrystalline synthetic permeation shows a strong influence of segregation on output flux magnitude. This approach is able to simulate also the short-circuit diffusion phenomenon. The comparison between different grain sizes and grain boundary thicknesses by means of the fitted apparent diffusivity shows the relationships between the registered flux and the characteristic parameters of traps.

**KEYWORDS:** Hydrogen diffusion, Permeation test, Finite Element Simulation, Grain Boundary trapping

## 1. INTRODUCTION

Numerous efforts have been put on the characterisation of metals and alloys behaviour in hydrogen environments. Microstructural phenomena operating during hydrogen embrittlement failures are still not entirely understood; however, it is accepted that damage depends on hydrogen local concentration in the Fracture Process Zone (Hirth 1980; Gerberich et al. 1996). Therefore, hydrogen transport, i.e. hydrogen entry, diffusion and trapping, determines the kinetics of crack initiation and propagation during the Hydrogen Assisted Cracking process (Turnbull 1993).

Modelling hydrogen-assisted fracture requires the implementation of a coupled scheme to simultaneously solve deformation, diffusion and damage equations (Martínez-Pañeda et al. 2018). Deformation problem includes plasticity formulations that could be modified considering hydrogen effects, e.g. hydrogen-enhanced localised plasticity (HELP) (Miresmaeili et al. 2010). Diffusion problem might be solved assuming different simplifications, from the simplest fickian equations to the most complex multi-trapping models including kinetic expressions and the influence of stress and plastic strain on hydrogen transport (Dadfarnia et al. 2011; Díaz et al. 2016a). Both deformation and hydrogen concentration fields directly influence crack propagation within the coupled finite element scheme; different damage models have been used to simulate hydrogen-assisted fracture (Olden et al. 2008; del Busto et al. 2017; Martínez-Pañeda et al. 2018), especially those assuming a decohesion mechanism in which the local fracture energy is reduced by hydrogen. Therefore, a fundamental aspect in modelling hydrogen-assisted fracture is the characterisation of trapping parameters so the diffusion equations are physically-based and local concentrations as realistic as possible.

Electrochemical permeation is one of the most used testing methods for characterising diffusion and trapping phenomena in metals and alloys. This technique requires two cells separated by the tested metallic sheet. In one cell, hydrogen is produced through a cathodic reduction while adsorption/absorption reactions take place in the entry surface of the specimen (Devanathan et al. 1963). After permeating, hydrogen output flux is registered in the oxidation cell. The influence of the microstructure has been extensively analysed, and it has been demonstrated how crystal defects delay permeation (Frappart et al. 2012). Thus, permeation trough a sample with a given number of imperfections, e.g. dislocations, grain boundaries, inclusions, etc., shows a lower diffusivity than a free-defect ideal lattice. This phenomenon has been described as hydrogen trapping and it is fundamental in the prediction and mitigation of hydrogen accumulation near a crack tip.

Permeation test is standardised by the standards ISO 17081:2014 and ASTM G148-97(2018); using these procedures only apparent diffusivity and apparent concentration in the entry surfaces might be obtained but this usually gives little information about the microstructural features of the material. Finite element simulations considering Fick's laws modified by trapping phenomena have been carried out by many researchers with the aim of elucidating the effect of microstructural parameters on permeation transients (Turnbull 2015; Raina et al. 2017; Vecchi et al. 2018a). Most of these works assume a modified mass balance including hydrogen concentration in trapping sites and following the pioneering work of Sofronis and McMeeking (1989). This approach, identified here as Continuum model, is revisited with the aim of evaluating the influence of binding energy and trap density. The effect of input concentration on the entry side is also studied within this framework. Additionally, the relationship between these three magnitudes (binding energy, trap density and input concentration) determines whether hydrogen permeation response deviates from the classical Fickian diffusion or not. This deviation is studied following the work of Raina et al. (Raina et al. 2017) in which three regimes are defined for hydrogen permeation.

Even though this approach is demonstrated to be consistent in hydrogen permeation modelling, the evaluation of grain boundary trapping is limited. Grain boundaries might be regarded as critical trapping sites: many authors have found hydrogen segregation in grain boundaries for nickel (Oudriss et al. 2012) and for iron (Ono and Meshii 1992). Additionally, there is a lack of consensus regarding the influence that accelerated or "short-circuit diffusion" of hydrogen through grain boundaries might have; atomistic simulations can give an insight of the competition between trapping and acceleration phenomena (Oudriss et al. 2012), but this fact must be empirically elucidated. It must be taken into account that the polycrystalline characteristics, e.g. Coincidence Site Lattice (CSL) and random boundaries distribution, misorientation angles, segregation of impurities, etc., determine the occurrence of one or another phenomenon.

A synergistic effect between impurity segregation, e.g. manganese, silicon or sulphur, might be found in hydrogen-related intergranular failures of steels (McMahon 2001). In these cases, fracture occurs before macroscopic yielding, so prognosis of intergranular embrittlement is fundamental for industry.

Intergranular fracture has been observed also in nickel alloys where the influence of grain boundary misorientation in hydrogen transport has been usually addressed (Oudriss et al. 2012). It has been demonstrated that grain-boundary engineering, in particular the control of special and random fractions of grain boundaries, might mitigate hydrogen embrittlement (Bechtle et al. 2009). The role of plasticity in intergranular fractures due to hydrogen in nickel has been revisited by (Martin et al. 2012).

Hydrogen embrittlement of martensitic steels has attracted an increasing attention since they have been proposed as materials for high pressure storage (Macadre et al. 2011) and are used in automotive industry (Venezuela et al. 2018). However, intergranular fracture is hard to be modelled due to the competition between fracture along prior austenite grain boundaries and fracture along lath martensite boundaries (Nagao et al. 2012). Due to the fracture surface appearance, the latter phenomenon is usually termed as "quasi-cleavage" despite its intergranular nature, i.e. propagation along martensite block and packet interphases (Álvarez et al. 2019). Novak et al. (2010) proposed a complex micro-mechanism for hydrogen-induced intergranular fracture of a 4340 high strength steel. These authors modelled fracture considering that hydrogen promotes dislocation pile-up impinging. Hence, decohesion of carbide/matrix interface triggers intergranular fracture.

Numerical modelling of hydrogen transport is usually focused on continuum approaches with the modified Fick's laws, i.e. a flux and a mass balance, as starting point – see, e.g. (Díaz et al. 2016a; Martínez-Pañeda et al. 2016a; del Busto et al. 2017). However, the continuum framework cannot evaluate geometrical factors, percolation phenomena or grain boundary connectivity (Hoch et al. 2015). Simulations following the work of Sofronis and McMeeking (1989) consider that traps are isolated, i.e. that flux between traps is negligible, which seems to be unlikely for grain boundary trapping sites. Few attempts have been made to explicitly reproduce the grain boundary trapping effect using a polycrystalline synthetic structure still considering continuum formulations (Hoch et al. 2015; Jothi et al. 2015). In the present work, a polycrystalline model is considered using a synthetic structure generated with a Voronoi tessellation. Generating different geometries, the grain size influence and the effect of grain boundary thickness are evaluated. A simplified model is considered in which the conventional Fick's laws govern hydrogen transport. The trapping behaviour is not taken into account in the mass balance for each material point but in the definition of two different materials for the grain and for the grain boundary with their corresponding diffusivity and solubility values. Thus, two approaches are considered and compared: a continuum 1D model and a polycrystalline model. Despite the nomenclature, both approaches are based on a continuum formulation, i.e. on a finite element framework in which a local mass balance is solved. Nevertheless, since the polycrystal reproduces explicitly the grain boundary diffusion barriers, permeation modelling relies on a two-scale approach.

After the model definition, the fitting procedure of the output flux in the simulated permeation test is presented. An apparent diffusivity is thus considered as a macroscopic parameter that measures the motion of hydrogen through the whole polycrystalline structure of 1-mm thickness. Results are then analysed and the relationship between flux evolution, apparent diffusivity and microstructural parameters is explained.

## 2. CONTINUUM 1D MODEL

Mass balance is the partial differential equation (PDE) governing hydrogen transport; from the traditional Fick's laws, some modifications might be introduced, particularly the introduction of additional terms that take into account trapping effects.

- **Selection of variables**: taking the nomenclature from Toribio and Kharin (2015) hydrogen modelling might consider a one-level system and only one concentration ($C$), a two-level system in which ideal diffusion is governed by lattice hydrogen and only one type of trap is considered ($C_L$ and $C_T$) or a generalised model with multiple trap types ($C_L$ and more than one kind of trap $C_{T,i}$). Sometimes, occupancy is the dependent variable rather than the concentration. In the one-level approach, trapping effects can be simulated by considering modified values of diffusivity and solubility whereas the two-level model considers explicitly hydrogen concentration in trapping sites.

$$\frac{\partial C_T}{\partial t} + \frac{\partial C_L}{\partial t} + \nabla \cdot \mathbf{j} = 0 \qquad (1)$$

- **Flux expression**: it is usually assumed that flux is proportional to lattice diffusivity, $D_L$, and to the gradient of hydrogen concentration in lattice sites. However, this assumption is only verified for low occupancy, for isolated traps or for low trap density. The accuracy of flux expression is usually overlooked in works dealing with hydrogen transport modelling (Díaz et al. 2016a).

$$\mathbf{j} = -D_L \nabla C_L \qquad (2)$$

- **Relationship between concentrations**. If hydrogen concentration in trapping sites is included in the mass balance as dependent variable, an additional equation relating $C_T$ (or each $C_{T,i}$) and $C_L$ must be considered. Thermodynamic equilibrium, as proposed by Oriani (1970), is usually assumed giving a univocal relationship that might be easily implemented in finite element (FE) codes. However, for some conditions this equilibrium is not fulfilled, and a kinetic formulation should be considered. In the latter case, McNabb and Foster's equation (McNabb and Foster 1963) is implemented to calculate hydrogen concentration in trapping sites.

In results presented in Section 5.1., two variables are implemented: $C_L$ and $C_T$, and only hydrogen flux between lattice sites is considered. Oriani's equilibrium is assumed so hydrogen concentration in trapping sites at each permeation distance is calculated following expression (3):

$$C_T = \frac{N_T}{1 + \frac{N_L}{C_L \exp\left(\frac{E_b}{RT}\right)}} \quad (3)$$

where $N_T$ is the trap density, $N_L$ the number of lattice sites per unit volume, $E_b$ the binding energy characterising the considered defect, $R$ the ideal constant of gases and $T$ the temperature. Operating expression (3) to obtain $\partial C_T/\partial t$ and rearranging mass balance in (1), an effective diffusivity might be thus defined as (Sofronis and McMeeking 1989):

$$D_{eff} = \frac{D_L}{1 + \frac{C_T}{C_L}(1 - \theta_T)} \quad (4)$$

Effective diffusivity is a local parameter that is defined through the mass balance modification expressed in (1) and assuming thermodynamic equilibrium and low lattice occupancy. This magnitude must not be confused in the present work with the apparent diffusivity $D_{app}$ obtained through the permeation transient fitting even though some works swap the use of both terms. Kharin (2014) highlights the confusion around this diffusivities and defines the term $D_{eff}$ expressed in (4) as an "operational diffusivity" since it is a mathematical rearrangement rather than a physical feature. Lattice diffusivity is considered as a theoretical value from first principle calculations for bcc iron (Jiang and Carter 2004) with $D_{L,0}$= 0.15 mm$^2$/s and $E_a$ = 8.49 kJ/mol (0.088 eV) considering an Arrhenius expression:

$$D_L = D_{L,0} \exp(-E_a/RT) \quad (5)$$

which gives a $D_L$ = 4987.5 µm$^2$/s at room temperature. The number of lattice sites is taken from (Hirth 1980) assuming tetrahedral occupation: $N_L$ = 5.09×10$^{29}$ sites/m$^3$. Even tough different values for $D_L$ for bcc iron have been found and $N_L$ depends on the preferential site and the number of hydrogen atoms that a site might allocate, the order of magnitude is always very similar. However, the trap density $N_T$ has been experimentally found in a range from 10$^{20}$ (Kumnick and Johnson 1980) to 10$^{27}$ (Oudriss et al. 2012) traps/m$^3$. Figure 1 represents trapping densities considered in different works (Kumnick and Johnson 1980; Sofronis et al. 2001; Juilfs 2002) dealing with hydrogen transport in bcc iron where the number of traps depends on plastic strain because dislocations are created. Oudriss et al. (2012) found trap densities around 10$^{25}$ to 10$^{27}$ traps/m$^3$ for grain boundary trapping in nickel, depending on the grain size and grain boundary nature. In this case, $N_T$ is associated to the density of Geometrically Necessary Dislocations (GND) (Ashby 1970; Martínez-Pañeda et al. 2016b).

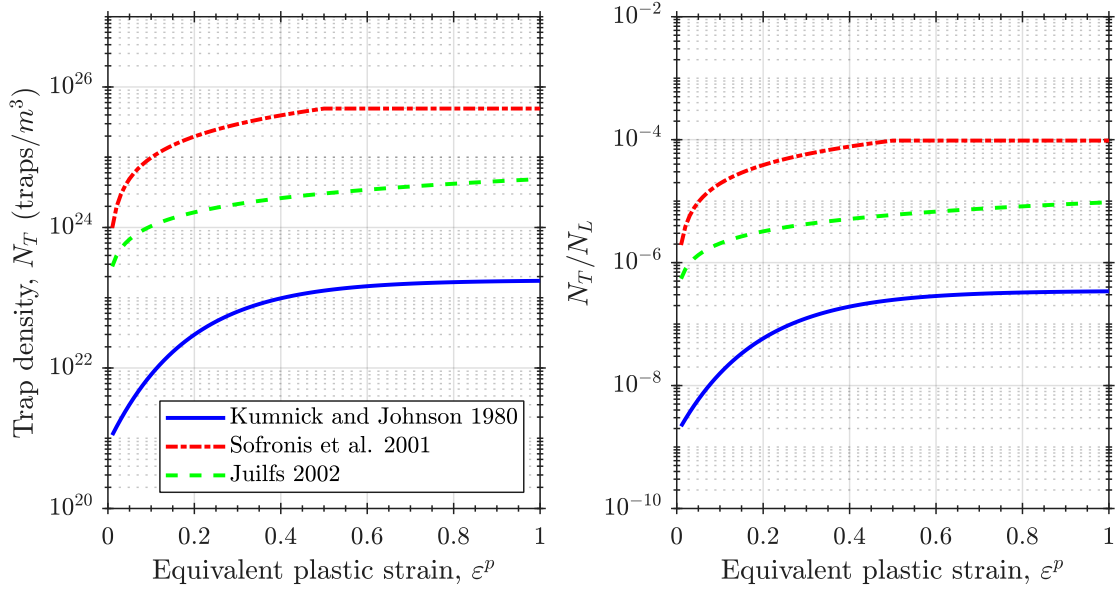

*Figure 1. Trap density as a funcion of equivalent plastic strain following the references* (Kumnick and Johnson 1980; Sofronis et al. 2001; Juilfs 2002)*; (a) $N_T$ expressed in sites/m³ and (b) normalised trap density $N_T/N_L$.*

The binding energy, as estimated from Thermal Desorption Analysis (TDA), for different traps takes values from 20 to 70 kJ/mol (Song 2015). This variation is critical since $E_b$, as seen in (3), is placed inside an exponential term. Thus, in the following simulated permeation tests a parametric study is carried out over the expected range of trap characteristic values. Simulated values are shown in Table 1.

| Relative number of trapping sites: $N_T/N_L$ | $10^{-10}$; $10^{-8}$; $10^{-6}$; $10^{-4}$; $10^{-2}$ |
|---|---|
| Binding energy [kJ/mol]: $E_b$ | 30;   45;   60 |

*Table 1. Parameters for analysing trapping effects using the continuum approach.*

For the continuum 1D simulation, COMSOL Multiphysics software has been used. The slab length is simulated considering the usual dimension of a permeation specimen for the standardised test, i.e. $L$ = 1 mm.

## 3. POLYCRYSTALLINE MODEL

Synthetic polycrystalline geometries are usually modelled using the mathematical procedure known as Voronoi tessellation. Initially, that partitioning requires the definition of a certain number of centroids, $c_i$, with random coordinates. The minimum distance between two centroids is constrained as the minimum diameter of the subsequently generated grains. From these centroids, the partition is performed imposing the following condition: a point $p_i$ belongs to the polygon corresponding to the $c_i$ centroid only if expression (1) is verified:

$$\|p_i - c_i\| < \|p_i - c_j\| \tag{6}$$

For each $j \neq i$. That condition implies that every point belonging to a grain is always closer to the centroid $c_i$ than to another centroid. In the present work, a python script is written for the finite element software ABAQUS CAE which is able to set the number of Voronoi polygons in a 2D geometry and the minimum distance between centroids. Obtained polygons are assumed to be representative enough of the real grain shape of a polycrystalline iron.

The permeation sample is considered as a slab with 1.0-mm thickness and 0.5-mm width. The tessellated 2D geometry introduces a tortuosity influence on permeation. In the 1.0×0.5-mm² slab, three synthetic polycrystals with 50, 200 and 800 grains are constructed. Even though the imposed minimum distance between each pair $c_i$ and $c_j$, represents the minimum grain diameter, average grain size is calculated after tessellation following the Intercept procedure specified by the ASTM E112-13 Standard. Table 2 shows the intercepted number of grains, averaged from the upper and lower lines in the models (Figure 2), which are translated into three average grain diameters.

Theoretically, grain boundaries are 2D interfaces, so the definition of grain boundary thickness might be regarded as physically-inconsistent. However, hydrogen trapping or segregation is demonstrated to occur in a finite strip around the boundary plane between two grains (Hoch et al. 2015). Additionally, trapping at grain boundaries has been attributed to the role of Geometrically Necessary Dislocations (GNDs) accumulated in those interfaces (Oudriss et al. 2012). Impurity segregation, e.g. in sensitised steels, might also play a role in hydrogen trapping. Considering all those phenomena, grain boundary thickness is modelled from the nanoscopic to the microscopic range (10, 100 and 1000 nm).

| | | | |
|---|---|---|---|
| Number of grains in the generated 1.0×0.5-mm² slab | 50 | 200 | 800 |
| Number of intercepted grains in a 1.0-mm test line | 7.0 | 19.0 | 36.5 |
| Average diameter $\bar{d}$ (µm) | 161.4 | 59.1 | 30.9 |

*Table 2. Evaluated grain sizes.*

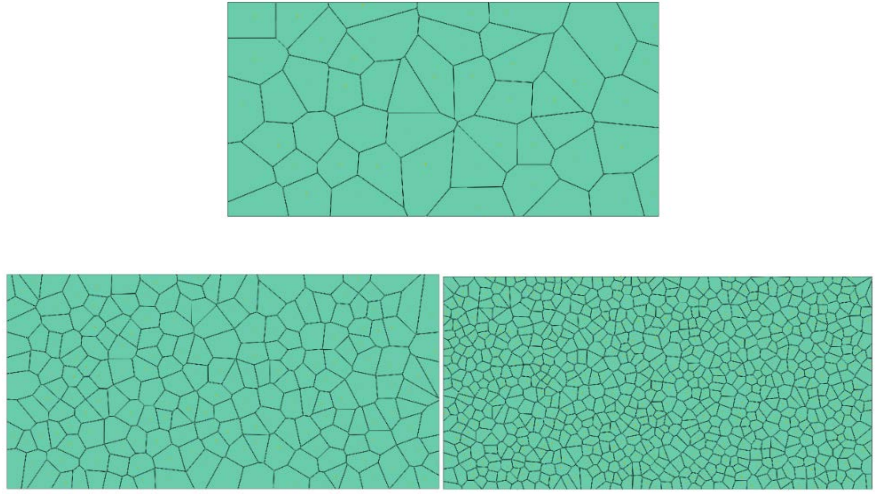

*Figure 2. Generated polycrystalline structures of 50, 200 and 800 grains.*

As previously mentioned, two materials are defined with different diffusivities with the aim of capturing hydrogen delay, or acceleration, in grain boundaries. Even though a crystal scale is considered, only isotropic diffusion is studied for the sake of simplicity. It must be noted that anisotropic effects in grain boundaries acting as diffusion barriers could have great influence. Within this numerical framework, mass balance governs hydrogen transport in each region so two PDE must be implemented with the corresponding parameters:

$$\frac{\partial C_L}{\partial t} - D_L \nabla^2 C_L = 0 \qquad (7)$$

$$\frac{\partial C_{gb}}{\partial t} - D_{gb} \nabla^2 C_{gb} = 0 \qquad (8)$$

where $C_L$ is the hydrogen concentration within grains, $D_L$ the ideal diffusivity, $C_{gb}$ the hydrogen concentration in grain boundaries and $D_{gb}$ a diffusivity determined by hydrogen jumps between trapping sites in grain boundaries. More details about the relationship between potential energy landscapes, hydrogen jumps and diffusivity values might be found in (Hoch 2015; Hoch et al. 2015)

The influence of grain boundary diffusivity is studied over a range for different $D_{gb}/D_L$ ratios, as shown in Table 3. For relative diffusivities of $10^{-4}$ and $10^{-2}$ a delaying effect is expected whereas the positive value $10^2$ might indicate diffusion acceleration due to grain boundary connectivity. Additionally, due to the wide range of binding energies associated with grain boundaries, segregation effects are assessed. A segregation factor can be defined as (Jothi et al. 2015):

$$s_{gb} = \frac{C_{gb}}{C_L} \qquad (9)$$

where the relationship between hydrogen concentration in grain boundaries $C_{gb}$ and the concentration in the adjacent lattice sites $C_L$ depends on the binding energy $E_b$. Thermodynamic equilibrium is assumed here so hydrogen concentration in grain boundaries should follow the expression (3) for $C_T = C_{gb}$. The evolution of segregation, i.e. of $C_T/C_L$, for a wide $C_L$ range is plotted in Figure 3 following expression (3). However, simulations for the polycrystalline model have been performed assuming a constant segregation factor. Three segregation factors are studied, as shown in Table 3.

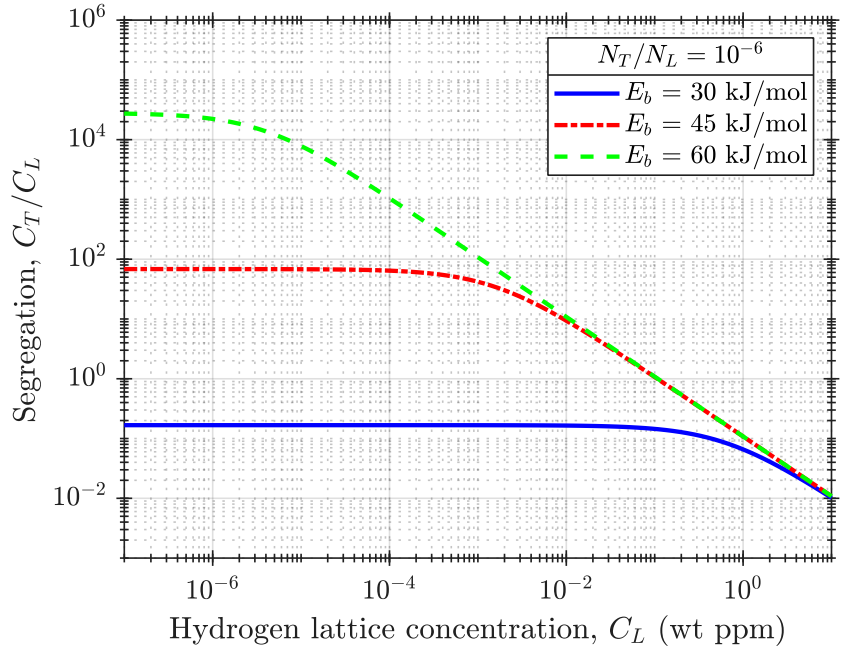

*Figure 3. Segregation, defined as the ratio $C_T/C_L$, as a function of hydrogen lattice concentrations.*

| Relative diffusivities: $D_{gb}/D_L$ | $10^{-4}$; $10^{-2}$; 1; $10^2$ |
|---|---|
| Segregation factor: $s_{gb}$ | 1; $10^2$; $10^4$ |

*Table 3. Microstructural parameters for analysing grain boundary effects.*

Due to the fact that Fick's laws are assumed within grains and along grain boundaries in the polycrystalline model, permeation behaviour is independent on concentration magnitude; only the steady state flux value varies with the input concentration. A constant concentration is fixed in the entry side as a boundary condition ($C_{in}$ = 1.0 wt ppm) whereas desorption is modelled in the output surface with $C_{out}$ = 0 wt ppm; the output flux obtained in the exit side is the analysed magnitude (Figure 4).

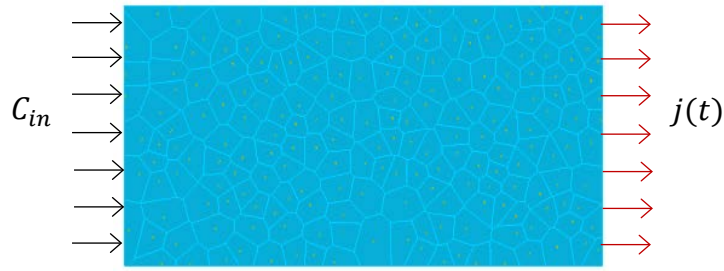

Figure 4. Constant concentration Boundary Condition and output flux.

## 5. FITTING OF OUTPUT FLUX

Calculation of diffusion parameters might be carried out following the ISO 17081:2014 Standard. Apparent diffusivity might be related to lag time or to breakthrough time. It must be noted that here, unlike in the Standards, effective diffusivity has been renamed as apparent diffusivity, since the former is usually considered as a microstructural characteristic magnitude and the latter an empirical value associated with the electrochemical permeation technique in this case. In the present work, lag time or breakthrough times are not used but the complete permeation transient is fitted into analytical expressions.

Analytical solutions of the Fick's second law are usually given for simple geometries – semi-infinite plates, thin plates – assuming homogeneous material, i.e. constant diffusivity independent of concentration too. When a constant concentration is imposed as a boundary condition, the analytical solution for diffusion in a thin plate might be expressed as ( ISO 17081:2014 Standard):

$$\frac{j(t)}{j_{ss}} = 1 + 2\sum_{n=1}^{\infty}(-1)^n \exp\left(-\frac{n^2\pi^2 D_{app}t}{L^2}\right) \qquad (10)$$

where $L$ is the slab thickness. Expression (8) represents Fourier's solution (ISO 17081:2014 Standard; Crank 1979; Turnbull et al. 1989) and it is used in the present work to fit the apparent diffusivity. However, it must be noted that Laplace's solution for a permeation is also usually considered (Turnbull et al. 1989; Frappart et al. 2010). Another option to find the apparent diffusivity is by simply finding the time at which flux reaches the 63 % of the steady state maximum, i.e. $j(t)/j_{ss} = 0.63$. The $t_{0.63}$ time, also called lag time, is related to the apparent diffusivity and the membrane thickness:

$$D_{app} = \frac{L^2}{6t_{0.63}} \qquad (11)$$

Steady state flux $j_{ss}$ is related to the constant concentration imposed at the entry side but $D_{app}$ should be independent of charging conditions under the hypothesis of being a material characteristic parameter. In the present simulations a constant concentration is imposed so the discussion of charging conditions is out of the scope of the work; but when empirical results are analysed, there is an uncertain condition in the entry surface that influences the numerical boundary conditions. Even though the analytical solution (10) is the only transient considered by the ISO 17081:2014 Standard, some authors have proposed that constant subsurface concentration is not always verified, especially in galvanostatic conditions. In that case, a constant flux is a more appropriate boundary condition (Montella 1999); kinetics of adsorption and absorption have been studied by some authors (Turnbull et al. 1996; Montella 1999; Turnbull 2015; Vecchi et al. 2018a, b) but discussion on boundary conditions is usually overlooked and should be better addressed in future research.

## 5. RESULTS

### 5.1. Continuum model

Within the two-level framework in which $C_L$ and $C_T$ are considered, the output flux in a FE permeation model is usually found in the exit side as:

$$j = -D_L \frac{\partial C_L}{\partial x}\bigg|_{x=L} \qquad (12)$$

And with a fixed concentration at the input side, $C_L(x=0) = C_{in}$, the steady state flux magnitude is independent of trapping phenomena (Raina et al. 2017):

$$j_{ss} = \frac{D_L C_{in}}{L} \qquad (13)$$

Legrand et al. (2014) placed effective diffusivity (4) inside the gradient operator so the authors conclude that steady state flux depends on trapping parameters (Bouhattate et al. 2011). However, this rearrangement is dubious (Kharin 2014) and here it is assumed that steady state flux does not depend on trapping parameters when exit flux follows expression (12). From this point of view, $j_{ss}$ should depend only on boundary conditions on the entry surface, i.e. on adsorption/absorption phenomena. However, it must be noted that when it is assumed that flux vector follows expression (11), fluxes from traps ($j_{TL}$), towards traps ($j_{LT}$) or between traps ($j_{TT}$) are neglected (Díaz et al. 2016a, b). This approximation is based on the assumption of deep potential wells or mutual remoteness of traps (Toribio and Kharin 2015). However, when modelling grain boundary trapping, it seems unlikely that traps are actually isolated even if $N_T \ll N_L$. If the conditions required for this assumption are not verified, the output flux at steady state would be influenced by trapping features.

Output flux obtained through FE simulations in a 1D model is plotted versus time for each of the analysed combinations of $E_b$ and $N_T/N_L$. Apparent diffusivity is obtained through a non-linear least-squares fitting algorithm in Matlab in which the expression (10) is implemented up to $n = 20$. Additionally, the $t_{0.63}$ time is also calculated. Numerically, steady state is defined when the output flux increment is less than $\Delta j/j < 10^{-8}$. This definition is important for the transient fitting.

These fitting operations are shown in Figure 5 for $C_{in}$ = 10⁻³ wt ppm, $E_b$ = 45 kJ/mol and two different trap densities. As expected, the higher trap density the lower apparent diffusivity is obtained. Even though the regression is satisfactory for both trap densities, it can be appreciated that for $N_T/N_L = 10^{-2}$, i.e. a very high trap density, the analytic transient flux is greater than the numerical flux before $t_{0.63}$ and is a little bit lower after that instant. This behaviour is confirmed in all simulations, so it is concluded that trapping effects produce a rise transient steeper than the flux predicted by Fick's laws.

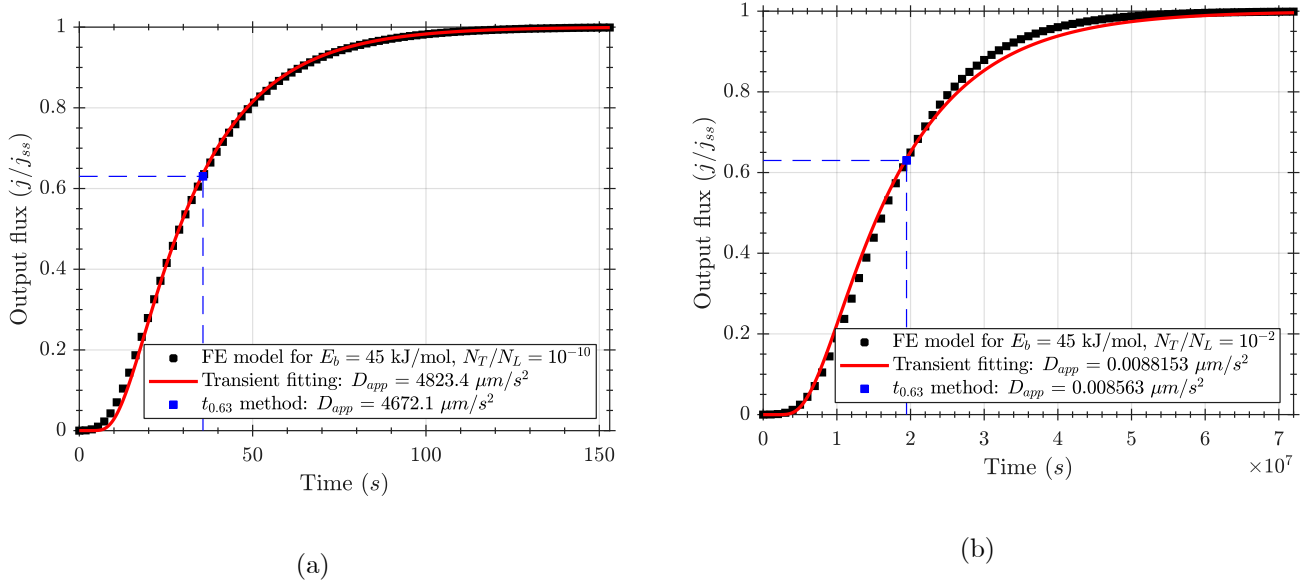

*Figure 5. Calculation of apparent diffusivity through transient fitting and $t_{0.63}$ method for $E_b$ = 45 kJ/mol, $C_{in}$ = $10^{-3}$ wt ppm and (a) $N_T/N_L = 10^{-10}$, (b) $N_T/N_L = 10^{-2}$.*

The effect of trapping phenomena might be evaluated following the approach proposed by Raina et al. (2017) through the definition of three regimes:

- Regime I: low trap occupancy due to low binding energy and low input lattice concentration. Trapping effects are minor and hydrogen permeation follows Fickian diffusion. Output flux can be modelled thus using the analytic rise transient (10) in which apparent diffusivity is fitted.
- Regime II: Deep trap limit (high binding energy) and high trap density.
- Regime IIIa: Deep trap limit (high binding energy), low trap density and low input lattice concentration.
- Regime IIIb: Deep trap limit (high binding energy), low trap density and high input lattice concentration.

By simulating hydrogen permeation through a material with a moderate trap density ($N_T/N_L$ = $10^{-6}$) and varying the binding energy and the input lattice concentration, three different permeation behaviours are found which can be explained using the defined regimes. As shown in Figure 6, for a binding energy equal to 45 kJ/mol and a limited hydrogen entry, a Fickian smooth permeation curve is found (Regime I). However, when strong traps are considered, i.e. $E_b$ = 60 kJ/mol, for the same input concentration, flux rise is delayed and becomes more abrupt so Regime II can be assumed. On the other hand, for $E_b$ = 45 kJ/mol an abrupt change is found if the concentration at the entry side is 1 wt ppm; in this case, the slope after the flux breaktrhough is more gradual which is typical of Regime IIIb, as found by Raina et al. (Raina et al. 2017).

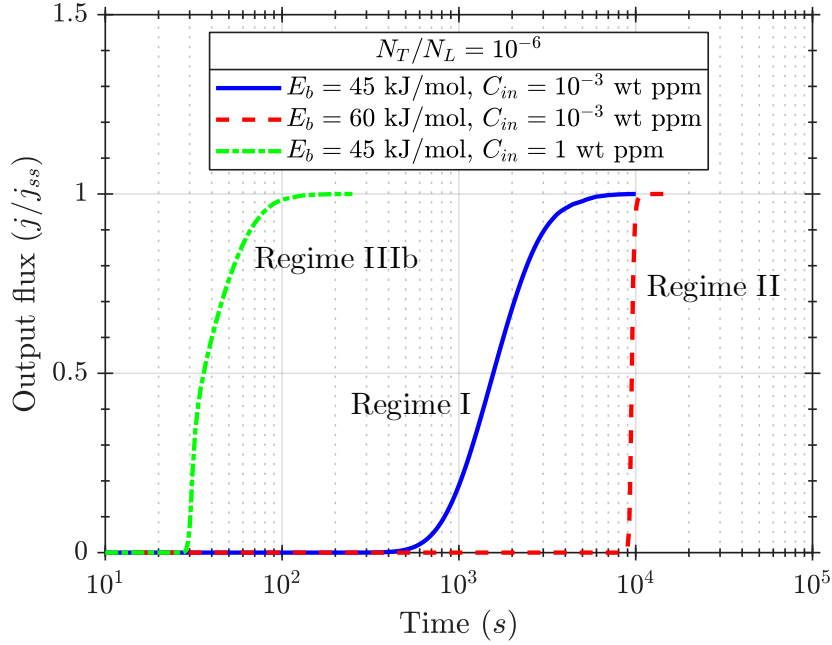

*Figure 6. Normalised output flux for different combinations of binding energies and input concentrations.*

Therefore, the relationships between $E_b$, $N_T$ and $C_{in}$ determine the permeation observed regime. This fact also influences Goodness of Fitting (GOF) of numerical permeation transients. Only for Regime I, i.e. when Fickian diffusion is predominant, the apparent diffusivity is fitted through a regression procedure considering the transient analytic solution. However, for Regimes II, IIIa or IIIb, the GOF is unacceptable and apparent diffusivity is calculated using the time lag method or, in this case, to the $t_{0.63}$ time. The cause underlying the steep rise flux in Regime II and IIIb might be better understood when the $C_L$ distribution is plotted along the thickness. Figure 7 represents the evolution of lattice concentration for low $C_{in}$ and different binding energies: (a) 30 kJ/mol, (b) 45 kJ/mol and (c) 60 kJ/mol. It can be seen that when strong traps are simulated (Figure 7.(c)), there is an advancing front that divides the filled and the empty region. Due to the high binding energy, hydrogen is not able to reach farther lattice sites until all the traps are filled. Thus, a bilinear $C_L$ distribution is obtained in Regime II and only when the front reaches $x = L$ the flux rises abruptly. Something similar is happening for $E_b$ = 45 kJ/mol with a high $C_{in}$ (Figure 7.(d)). In this case, lattice sites beyond the front are completely empty, but in this Regime the distribution is smoother.

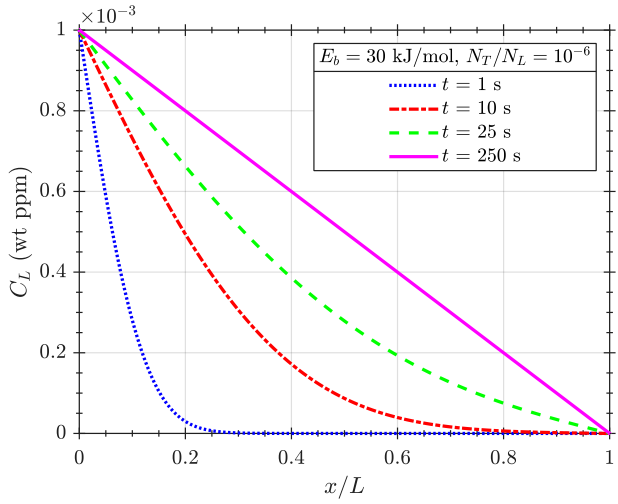

(a)

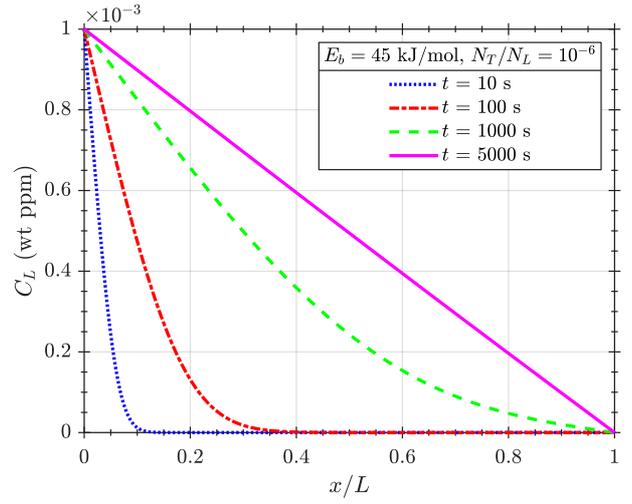

(b)

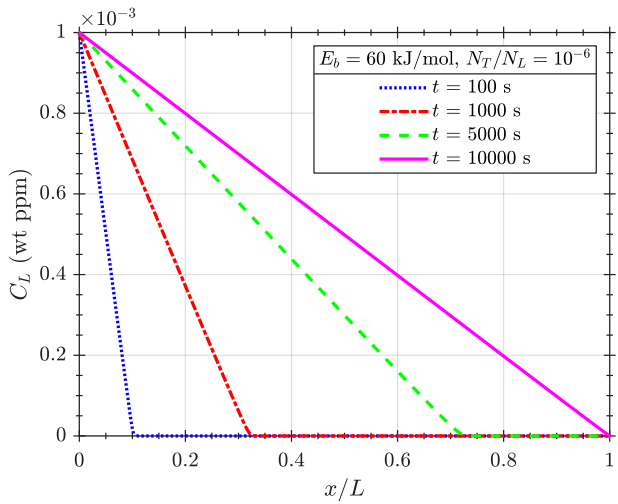

(c)

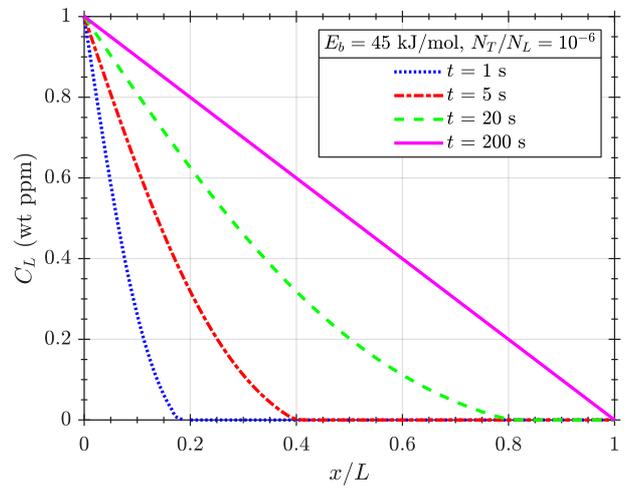

(d)

*Figure 7. Evolution of hydrogen concentration in lattice sites for $N_T/N_L = 10^{-6}$ and (a) $E_b = 30$ kJ/mol, (b) $E_b = 45$ kJ/mol, (c) $E_b = 60$ kJ/mol and $C_{in} = 10^{-3}$ wt ppm. (d) $E_b = 45$ kJ/mol and $C_{in} = 1$ wt ppm.*

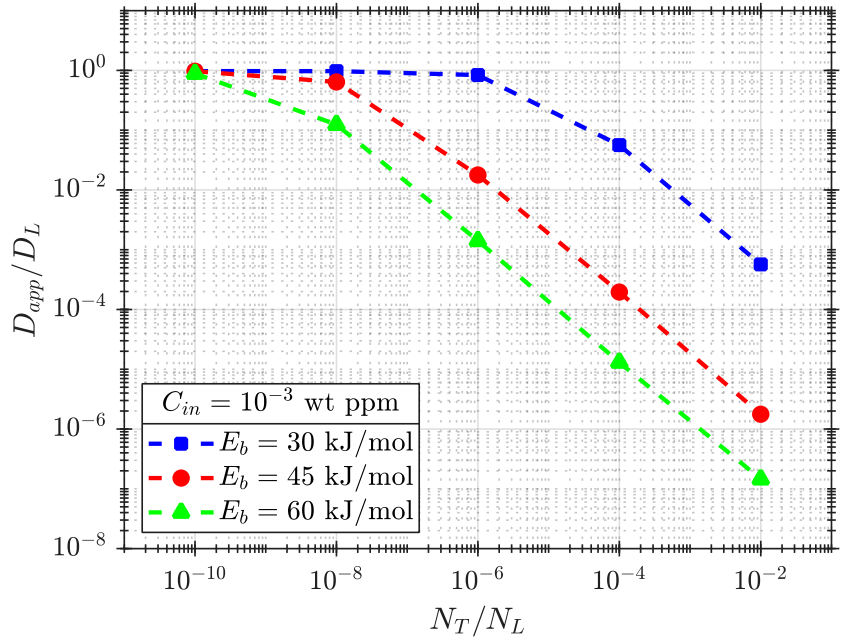

*Figure 8. Influence of binding energy and trapping density on apparent diffusivity for $C_{in}$ = 10⁻³ wt ppm.*

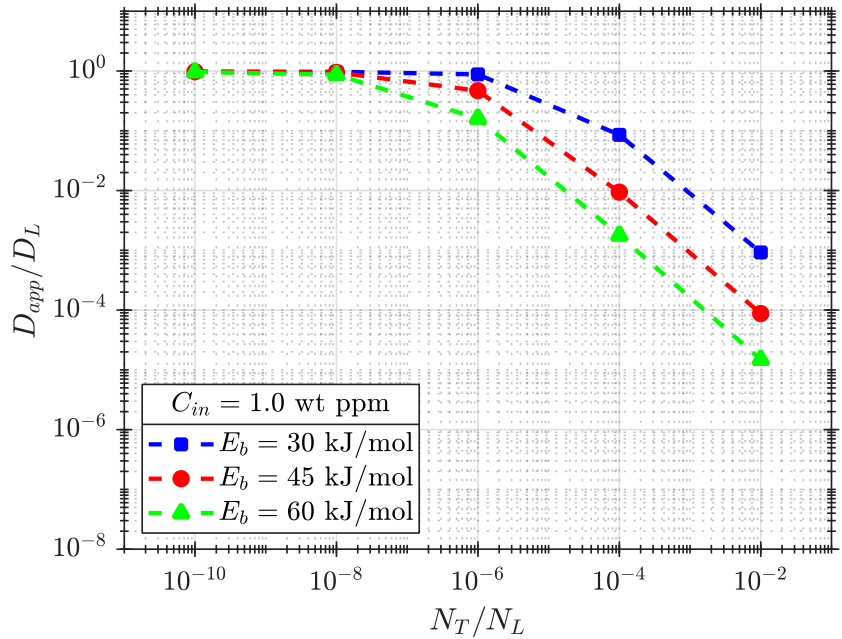

*Figure 9. Influence of binding energy and trapping density on apparent diffusivity for $C_{in}$ = 1 wt ppm.*

FE simulations are performed in COMSOL Multiphysics software over the range of parameters shown in Table 1. Apparent diffusivity has been fitted following Fourier's method, i.e. expression (10), in those cases where GOF is acceptable. However, for conditions lying on Regimes II or III the $D_{app}$ coefficients shown in Figures 8 and 9 where found using the lag time expression. Figures 8 and 9 might be linked with the results shown by Raina et al. (Raina et al. 2017) in Figures 6 and 7 where non-dimensional values are considered:

$\bar{N} = N_T/N_L$ and $\overline{\Delta H} = E_b/RT$, and lag time is proportional to apparent diffusivity. In the present paper, an exponential decay –since the x-axis is plotted in logarithmic scale– is found for apparent diffusivity at high trap densities (Figures 8 and 9). The curves corresponding to different hydrogen lattice concentration in the entry side show similar behaviour but a lower binding energy effect for $C_{in}$ = 1 wt ppm.

For the sake of illustration, permeation tests extracted from literature (Dietzel et al. 2006) for a low-alloy structural steel have been analysed in order to highlight the importance of the proposed regimes, the influence of trapping densities ($N_T$) and the effect of input concentration. These authors carried out permeation tests for different levels of deformation and calculated an apparent diffusivity using the lag time method; however, they identified this global value with the "operational" effective diffusivity as expressed in equation (4) and fitted the density of traps to a power law:

$$N_T = N_{T,0} + N_{T,1}\varepsilon_p^{0.7} \tag{14}$$

where equivalent plastic strain is expressed as a percentage, and the constants take the values: $N_{T,0}$= 8.8×10$^{22}$ traps/m³ and $N_{T,1}$= 4.8×10$^{24}$ traps/m³. Considering a concentration-independent effective diffusivity, the authors fitted the binding energy as $E_b$ = 42.1 kJ/mol. Dietzel et al. (2006) assume that hydrogen concentration in the entry side was the value obtained through the standard procedure from the steady state flux, giving $C_{L,0}$ = 2.1×10$^{22}$ atoms/m³ = 0.045 wt ppm. If these values of trapping density, binding energy and input hydrogen concentration are implemented in the 1D continuum model, results in Figure 10 are obtained, showing very poor agreement.

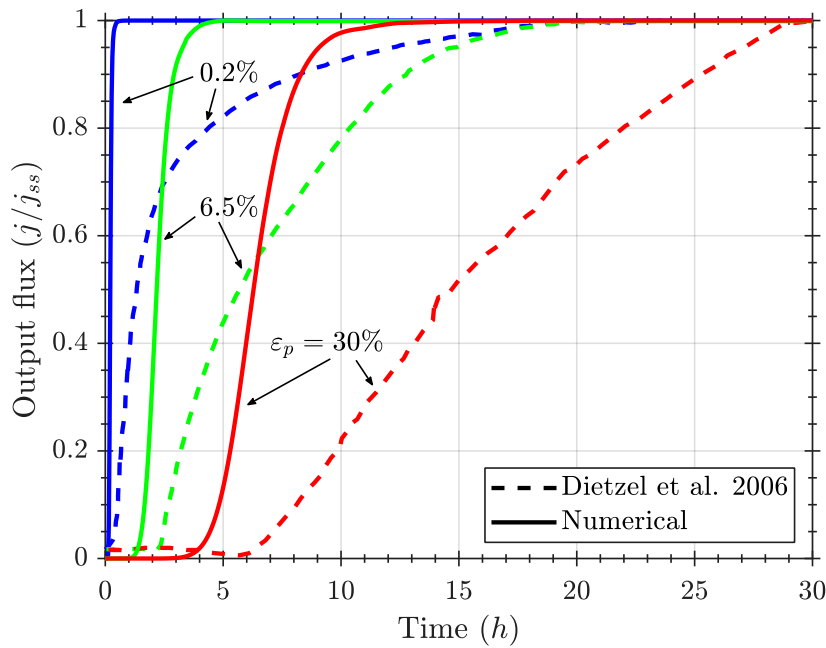

*Figure 10. Normalised output flux for different values of equivalent plastic strain extracted from (Dietzel et al. 2006) and simulated considering $C_{in}$= = 2.1×10²² atoms/m³ = 0.045 wt ppm.*

However, it has been previously shown how permeation transients are not concentration-independent and neither the apparent diffusivity. Hydrogen concentration at the entry side ($C_{in}$ = 2.1×10$^{22}$ atoms/m³ = 0.045 wt ppm) was calculated trough the steady state flux and the apparent diffusivity. If a lower entry concentration ($C_{in}$ = 0.01 wt ppm) is considerd, as plotted in Figure 11, experimental permeation tests are better fitted. It must be noted that trap densities deviate from the expression (14) for very low plastic deformations (Dietzel et al. 2006). Additionally, regression of $N_T - \varepsilon_p$ points assumed concentration-independent diffusivity which might be very inaccurate.

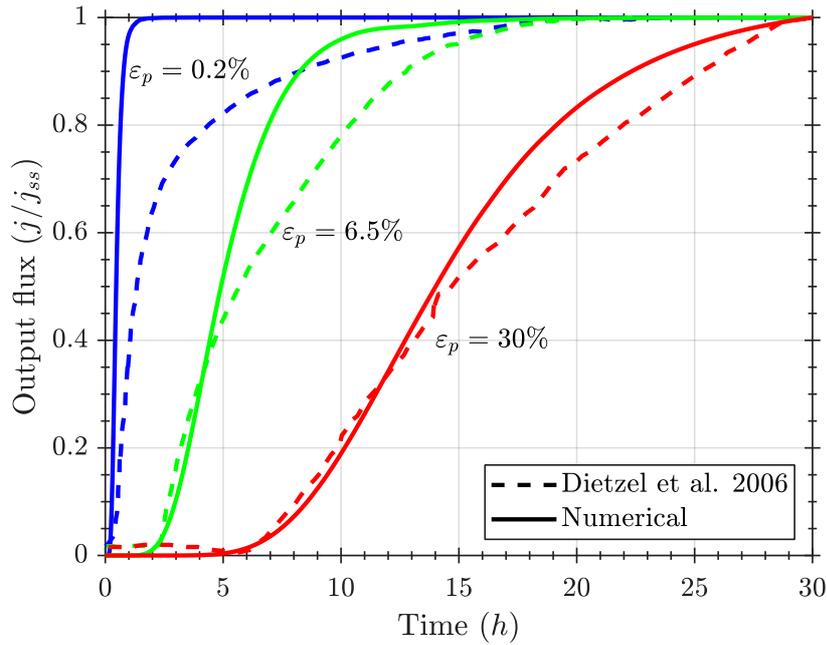

Figure 11. Normalised output flux for different values of equivalent plastic strain extracted from (Dietzel et al. 2006) and simulated considering $C_{in}$ = 0.01 wt ppm.

### 5.2. Polycrystalline model

Depending on the microstructural parameters, i.e. segregation factor and diffusivities, different concentration patterns are found during permeation simulations. In Figure 12, hydrogen distribution is show for a low grain boundary diffusivity and strong segregation.

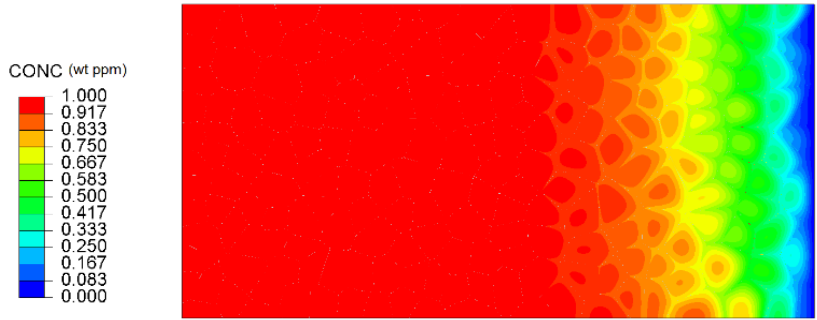

*Figure 12. Lattice hydrogen concentration distribution for $s_{gb}$ = 10⁴ and $D_{gb}/D_L$ = 10⁻⁴ at steady state.*

Since the concentration profiles are difficult to analyse, the output flux is studied and fitted in order to find apparent diffusivities as macroscopic indicators of trapping effects.

Hydrogen permeation trough a polycrystal shows that a low diffusivity along grain boundaries ($D_{gb}/D_L \ll 1$) promotes a delay in output flux and a lower steady state magnitude. On the other hand, when $D_{gb}/D_L \gg 1$ hydrogen exit is accelerated, and steady state flux is higher than the ideal lattice flux. Polycrystalline models are able, in contrast to the continuum model presented in the previous section, to simulate the trapping influence on hydrogen output flux. Figure 13 represents both situations, i.e. delay and acceleration due to a low or a high grain boundary diffusivity, respectively. These results are obtained without segregation, so hydrogen concentration is not enhanced in grain boundaries.

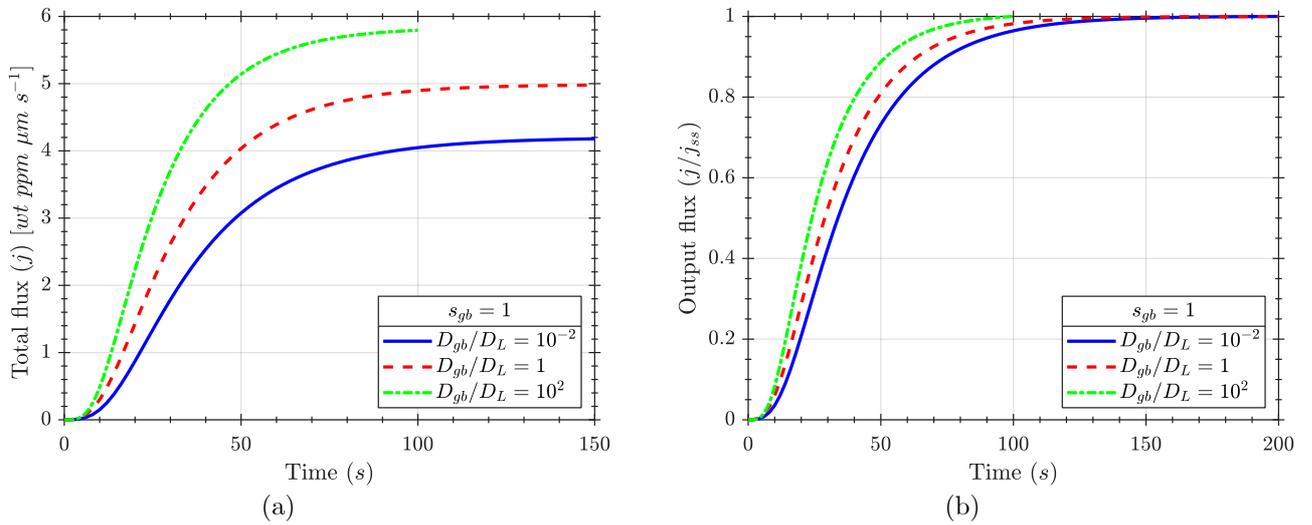

*Figure 13. Grain boundary diffusivity effect for 200 grains, $t_{gb}$ = 100 nm and $s_{gb}$ = 1.*

Even if hydrogen moves fast in grain boundaries, i.e. $D_{gb}/D_L$ = 1, a global delay is found in the permeation output flux when a segregation factor is considered (Figure 14). This can be explained because when hydrogen concentrations in grain boundaries are very high, grains tend to be depleted and the diffusion from the centre to the boundaries of a grain overweighs the macroscopic permeation.

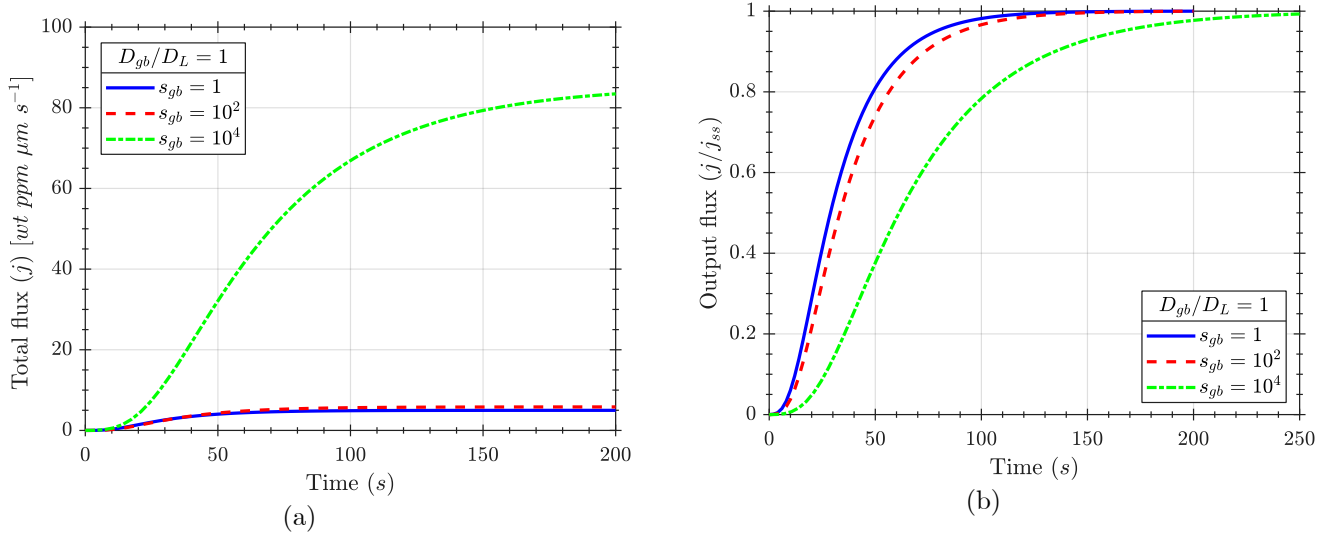

*Figure 14. Segregation effect for 200 grains, $t_{gb}$ = 100 nm and $D_{gb}/D_L$ = 1.*

For a significant delay in grain boundary diffusion, macroscopic delay is increased, as can be seen in Figure 15, in which time is plotted in logarithmic scale. However, a contradictory result is found for the segregation effect: strong trapping $s_{gb}$ = 10⁴ results in a permeation faster than for $s_{gb}$ = 10² or for no segregation.

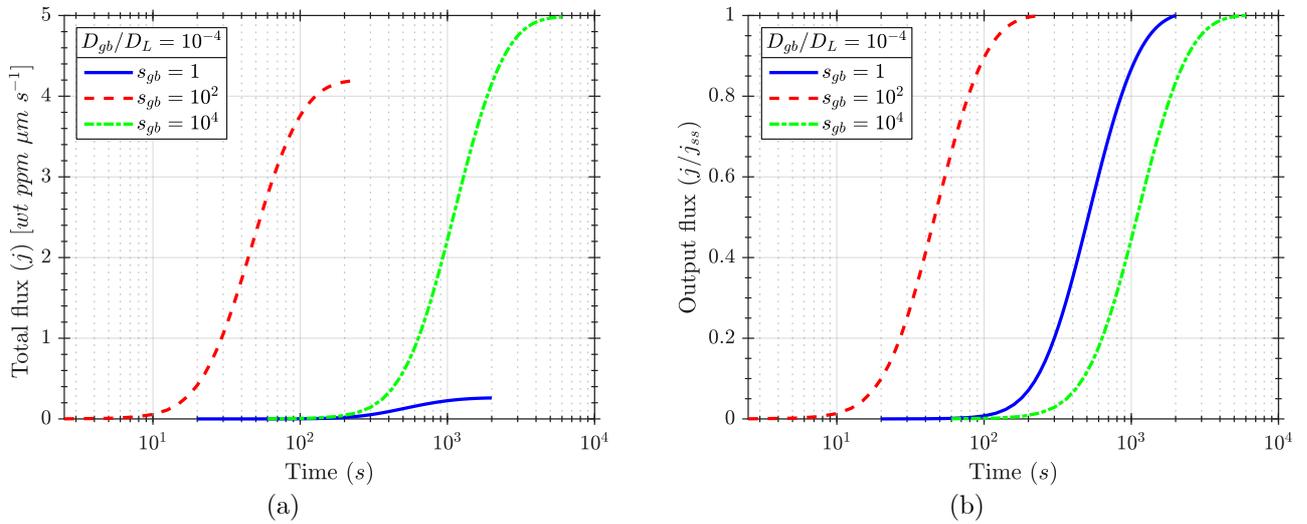

*Figure 15. Segregation effect for 200 grains, $t_{gb}$ = 100 nm and $D_{gb}/D_L$ = 10⁻⁴.*

In order to confirm the effect and to analyse the whole range of diffusivities and segregation factors, apparent diffusivities are determined by fitting the output fluxes.

Finite element results are fitted to the permeation transient flux expressed in (10) with a non-linear Least Squares algorithm. A fitting example is shown in Figure 16. A very good fitness is obtained for $D_{gb}/D_L$ < 1 ratios because the output flux is dominated by hydrogen desorption in grains whereas the fitting is worse for accelerated diffusion, i.e. for $D_{gb}/D_L$ = 10² since hydrogen exit from grain boundaries has an important weight in the output flux. Nevertheless, in both cases smooth curves are found that might be identified as Regime I, i.e. Fickian diffusion, within the framework previously presented.

The obtained $D_{app}$ for each pair of values $s_{gb}$ and $D_{gb}/D_L$ is plotted in Figure 17 for the structure with 200 grains and a grain boundary thickness of 100 nm. Apparent diffusivities are normalised using the lattice diffusivity value of $D_L$ = 4987.5 µm²/s.

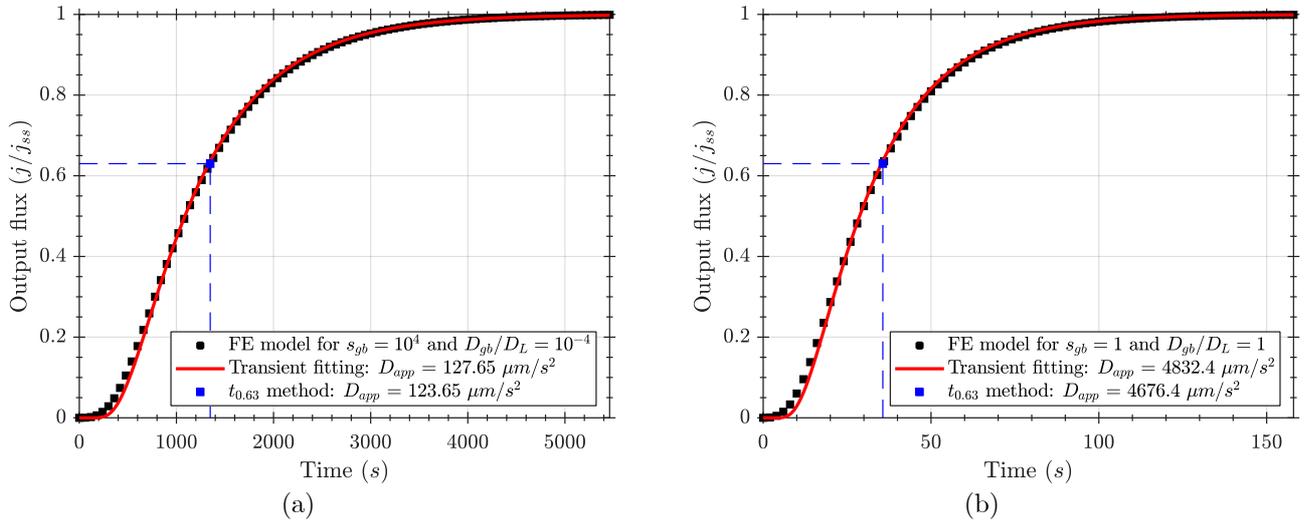

Figure 14. Fitting of FE results to permeation transient (4) for $s_{gb}$ = 10⁴ and $D_{gb}/D_L$ = 10⁻⁴ (200 grains and $t_{gb}$ = 100 nm)

As expected, for the pair of values $s_{gb}$ = 1 and $D_{gb}/D_L$ = 1, the apparent diffusivity coincides with the lattice coefficient, $D_{app}/D_L$ = 1, since no trapping effect is considered. The higher the grain boundary diffusivity, the higher the apparent diffusivity; however, this trend strongly depends on segregation. While the slope for $s_{gb}$ = 1 is very small and increases at low ratios $D_{gb}/D_L$, the trend is the opposite for $s_{gb}$ = 10² and $s_{gb}$ = 10⁴.

As shown in Figure 17, the result for very low diffusivities along grain boundaries ($D_{gb}/D_L$ = 10⁻⁴) seems contradictory: apparent diffusivity without segregation ($s_{gb}$ = 1) is higher than the value obtained for $s_{gb}$ = 10⁴, but lower than $D_{app}$ corresponding to $s_{gb}$ = 10². This fact was attributed to the effect of segregation on the output flux; for weak trapping, hydrogen transport is delayed but grain boundaries are not completely filled so total output flux is not affected by hydrogen exit from grain boundaries, whereas for strong traps diffusion is completely obstructed by grain boundaries near the entry side. Within an intermediate segregation range, hydrogen delay is reduced but a rise in the permeation transient might be registered early because of the enhanced hydrogen exit through grain boundaries. This phenomenon has been experimentally found using silver decoration techniques (Koyama et al. 2017b, a).

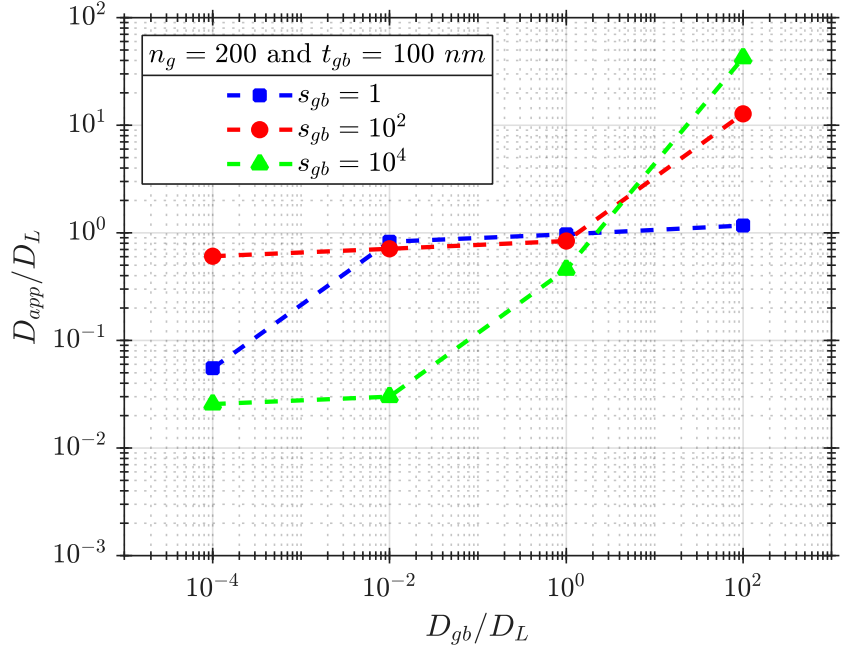

*Figure 17. Fitting results (200 grains, $t_{gb}$ = 100 nm)*

To confirm the segregation and diffusivity effects, the geometries shown in Figure 2, and whose parameters are collected in Table 3, are also analysed. For both the 50-grain and the 800-grain polycrystals, i.e. for the coarse-grained, (Figure 18.a.) and fine-grained (Figure 18.b) structures, respectively, a very similar tendencies are found. Nevertheless, the 50-grain simulated slab shows a lower change in apparent diffusivity due to grain boundary trapping. Both acceleration and delaying effects increase for the fine grain structure because a larger amount of grain boundaries exists. This expected influence is confirmed for the 800-grain crystal in which the $D_{app}$ variation over the $D_{gb}/D_L$ range is greater. For example, the lowest apparent diffusivity for 800 grains corresponding to $s_{gb}$ = 10$^4$ and $D_{gb}/D_L$ = 10$^{-4}$ gives a value of only $D_{app}$ = 62.3 µm²/s.

The main limitation of this formulation is that permeation fluxes and the corresponding fitted apparent diffusivities are independent of hydrogen concentration. However, it has been shown in the previous section that the continuum model, which includes a physically-based mass balance modification, predicts different regimes depending on the input concentration, i.e. for different charging conditions the obtained apparent diffusivity might be completely different. To be more realistic, the polycrystalline model could be modified to consider occupancies rather than total concentrations.

The 1D model can be related with grain boundary trapping and polycrystalline features when density of defects $N_T$, and binding energy $E_b$, represent grain boundary values $N_{T,gb}$ and $E_{b,gb}$. According to some references, (Song et al. 2013; Liu et al. 2019), trap densities associated with grain boundaries depend on burgers vector and average grain size:

$$\frac{N_{T,gb}}{N_L} = \frac{b}{d} \tag{15}$$

Burgers vector for bcc iron is 0.287 nm (Song et al. 2013), so for the considered microstructures the corresponding density of traps are show in Table 4

| Number of grains in the generated 1.0×0.5-mm² slab | 50 | 200 | 800 |
|---|---|---|---|
| Average diameter $\bar{d}$ (μm) | 161.4 | 59.1 | 30.9 |
| $N_{T,gb}/N_L$ | 1.78×10⁻⁶ | 4.86×10⁻⁶ | 9.29×10⁻⁶ |

*Table 4. Trap density calculated considering different average grain diameters of the simulated polycrystals.*

In a permeation test it is hard to isolate grain boundary effects since dislocations and impurities are always present within the grains. Nonetheless, the studied polycrystalline 2D geometries with grain sizes between 30.9 and 161.4 μm might be approximated by 1D models considering trapping energies about 10⁻⁶ times the density of lattice sites.

The influence of grain boundary thickness is analogous to the grain size effect; a very thin grain boundary ($t_{gb}$ = 10 nm) for 200 grains produces a very slight delay in diffusion, especially for $D_{gb}/D_L < 1$ ratios, as shown in Figure 16.c. Similarly, a very thick grain boundary ($t_{gb}$ = 1000 nm) results in an extreme variation in $D_{app}$ for all the segregation factors considered (Figure 18.d.). Both results confirm that the fraction of grain boundaries in comparison with the analysed permeation sample is a critical parameter in the apparent diffusivity that is empirically found.

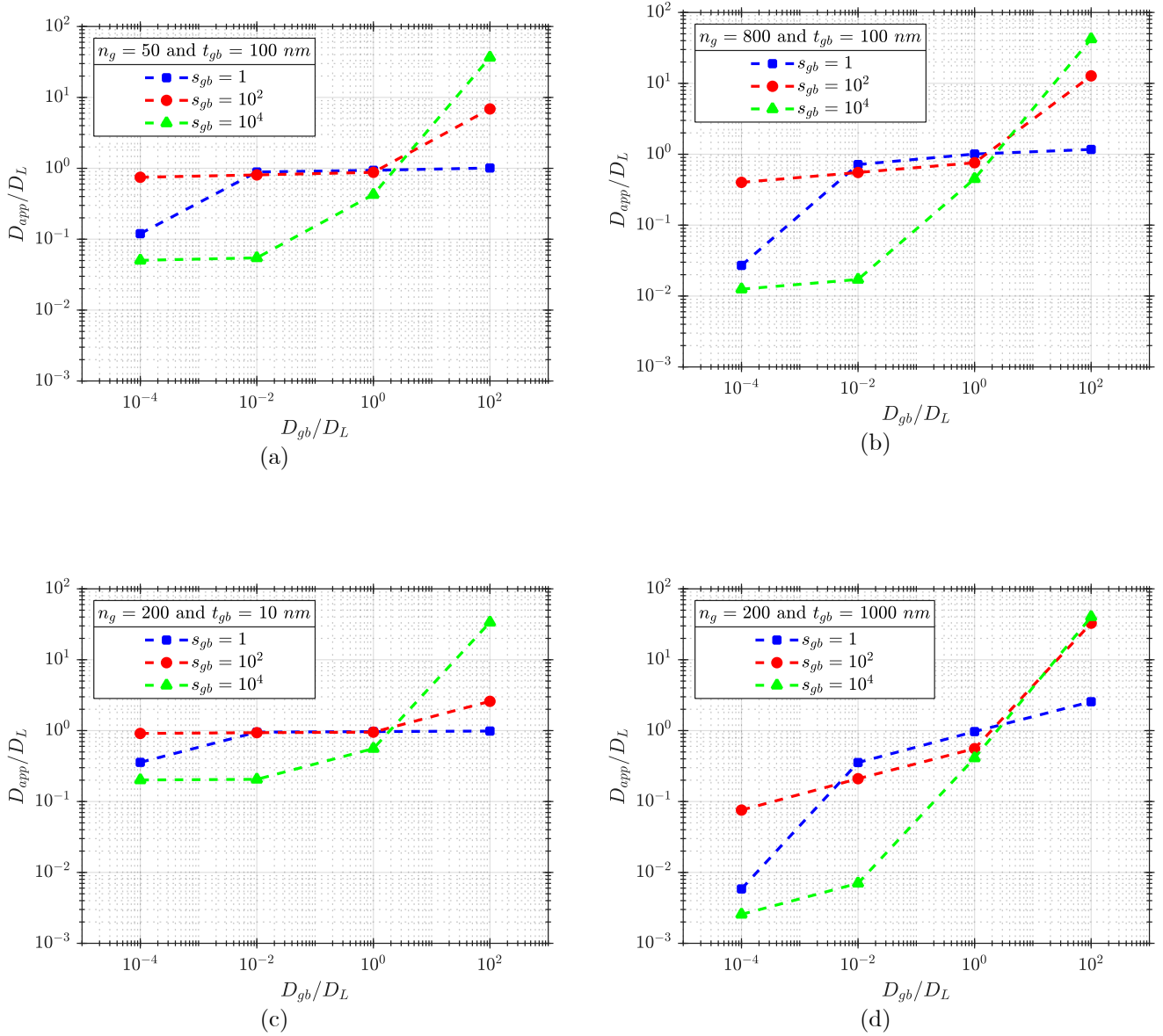

*Figure 18. Fitting results; a) 50 grains and $t_{gb}$ = 100 nm, b) 800 grains and $t_{gb}$ = 100 nm, c) 200 grains and $t_{gb}$ = 10 nm, d) 200 grains and $t_{gb}$ = 1000 nm.*

For the finer grain size, i.e. for 800 grains corresponding to an average diameter of 30.9 µm, it can be seen that apparent diffusivity might be substantially decreased. This numerical observation could be related to the beneficial effect of grain refinement in embrittlement mitigation that has been experimentally demonstrated in different works (Takasawa et al. 2012; Park et al. 2015; Macadre et al. 2015). Park et al. (2015), suggested that hydrogen segregation was higher in coarse grains compared to a fine-grained polycrystal due to the higher fraction of mechanical twins. However, total concentration measured by TDA was independent of grain size so when the same amount of hydrogen must be redistributed over a higher number of traps, i.e. in a fine-grained material, it is expected that local concentration in grain boundaries are lower. This effect was also confirmed by (Takasawa et al. 2012) who also pointed out that grain refinement mitigate embrittlement due to the reduction of the slip length of dislocations. Competition between dislocation-boundary interaction and hydrogen effects should be better understood. In this work, the retardation of diffusion in fine-grained materials has been numerically demonstrated for $D_{gb}/D_L < 1$; this fact might be critical in time-dependent fractures (i.e. strain rate dependence of hydrogen embrittlement). However, as found by other authors, (Takasawa et al. 2012; Park et al. 2015; Macadre et al. 2015), the grain refinement mitigation of embrittlement in Slow Strain Rate tests (SSRT) is expected to happen due to the redistribution of the same amount of hydrogen over a higher boundary surface so the local segregation and the subsequent intergranular decohesion is reduced.

It is hard to compare experimental results with the polycrystalline model predictions for pure iron because grain size effects on hydrogen permeation have been studied mainly for nickel alloys (Oudriss et al. 2012). Even though many works have dealt with the influence of different processes on hydrogen permeation in steels, e.g. heat treatment (Gesnouin et al. 2004; Lan et al. 2016) or work-hardening (Kumnick and Johnson 1980; Dietzel et al. 2006), microstructural features related to grain boundaries are not usually correlated. In the work of (Van den Eeckhout et al. 2017), hydrogen permeation through pure iron (Armco) is studied after different cold rolling levels. The as received pure iron corresponded to a grain size of 30 μm while the deformed samples showed elongated grains. Normalised permeation transients are represented in Figure 19.a. The polycrystalline results presented for the 800-grain model can be compared to the AR condition since grain sizes are equal (30 μm), as shown in Table 2, and the grain structure found by Van den Eeckhout et al. is isotropic. In that case, apparent diffusivity was fitted as 592 μm²/s, which is here normalised considering $D_L$ = 4987.5 μm²/s, i.e. $D_{app}/D_L$ = 0.119. Figure 19.b. illustrates how the relationship between segregation and grain boundary diffusivity should correspond to a point in the horizontal line $D_{app}/D_L$ = 0.119. In order to simulate permeation through anisotropic grain structures due to cold rolling, the synthetic grain generation should include a control algorithm reproducing elongation.

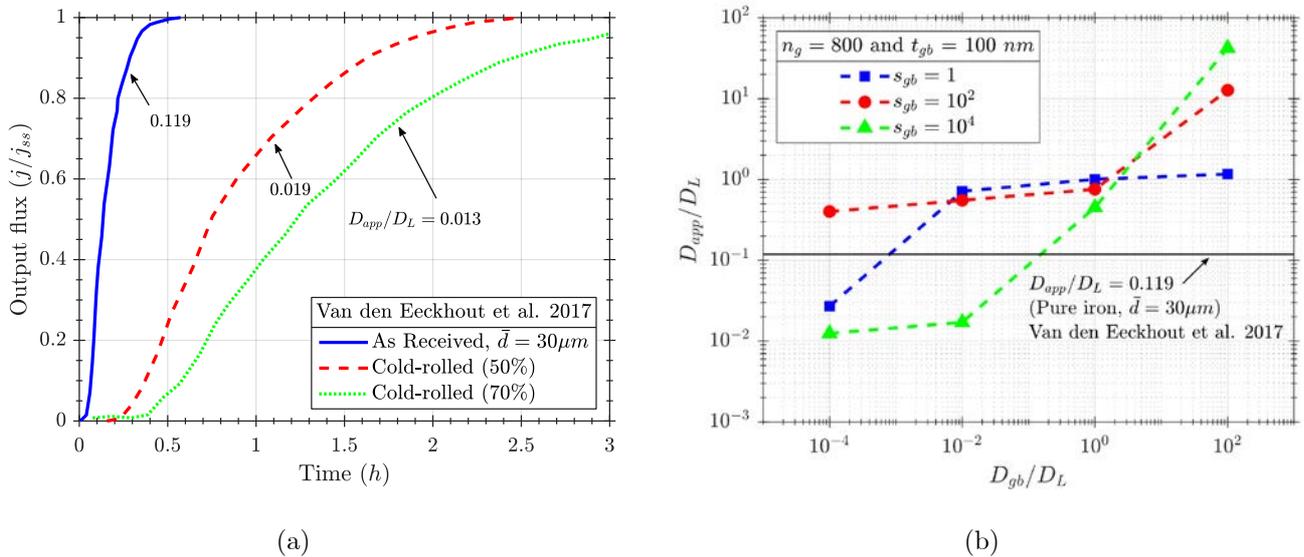

Figure 19. (a): experimental results of hydrogen permeation for pure iron and different thickness reductions after to cold-rolling (Van den Eeckhout et al. 2017). (b) Comparison of numerical results (parametric study for 800 grains) and experimental results for pure iron with $\bar{d}$ = 30 μm.

## 6. CONCLUSIONS

A methodology has been established for the simulation of hydrogen permeation through two approaches: (i) considering a 1D continuum model and (ii) generating a polycrystalline structure. The former is the most common framework since the work of Sofronis and McMeeking (1989) and relies on the modification of the mass balance including trapping influence. Even though it is useful for evaluating hydrogen transport, the concept of trap density is generic and might lack of physical meaning. On the other hand, the polycrystalline model is able to simulate a synthetic structure through a Voronoi tessellation and it is especially useful for the evaluation of trapping at grain boundaries. Microstructural parameters that could be evaluated through atomistic simulations are considered in a multiscale approach: lattice diffusivity, grain boundary diffusivity and segregations. Additionally, geometric features as grain size or grain boundary thickness are analysed as critical factors for the permeation output results. As expected, for very low grain boundary diffusivities the output flux is delayed so the apparent diffusivity decreases. However, the evaluation of a combined mechanism including segregation effects, i.e. a higher concentration in grain boundaries, is not straightforward and some coupled influences are hard to differentiate. This polycrystalline model must be related to continuum models based on a modified mass balance including density of trapping sites and binding energies. However, some simplifications made in the present work must be addressed in the future. For instance, the isotropy assumed here might be substituted considering a diffusivity tensor obtained through atomistic simulations. The difference between a 2D and a 3D permeation should also be evaluated using percolation theories and 3D finite element models. Additionally, the nature of grain boundaries, e.g. fraction of random boundaries or misorientation angles, influences hydrogen trapping and the possibility of short-circuit diffusion. Another possible implementation approach is based in Crystal Plasticity Finite Element Method (CPFEM) in which dislocation density might be coupled with trapping effects in a strained specimen. Future research will also focus on the influence of hydrogen entry processes, i.e. adsorption and absorption, and the permeation fluxes that are obtained considering realistic boundary conditions.

## ACKNOWLEDGMENTS

The authors gratefully acknowledge financial support from the project MINECO Refs: MAT2014-58738-C3-2-R and RTI2018-096070-B-C33. E. Martínez-Pañeda acknowledges financial support from the People Programme (Marie Curie Actions) of the European Union's Seventh Framework Programme (FP7/2007-2013) under REA grant agreement no. 609405 (COFUNDPostdocDTU).

## REFERENCES

Álvarez G, Peral LB, Rodríguez C, et al (2019) Hydrogen embrittlement of structural steels: Effect of the displacement rate on the fracture toughness of high-pressure hydrogen pre-charged samples. Int J Hydrogen Energy 44:15634–15643. doi: 10.1016/J.IJHYDENE.2019.03.279

Ashby MF (1970) The deformation of plastically non-homogeneous materials. Philos Mag A J Theor Exp Appl Phys 21:399–424. doi: 10.1080/14786437008238426

Bechtle S, Kumar M, Somerday BP, et al (2009) Grain-boundary engineering markedly reduces susceptibility to intergranular hydrogen embrittlement in metallic materials. Acta Mater 57:4148–4157. doi:

10.1016/J.ACTAMAT.2009.05.012

Bouhattate J, Legrand E, Feaugas X (2011) Computational analysis of geometrical factors affecting experimental data extracted from hydrogen permeation tests: I – Consequences of trapping. Int J Hydrogen Energy 36:12644–12652. doi: 10.1016/J.IJHYDENE.2011.06.143

Crank J (1979) The mathematics of diffusion. Oxford university press

Dadfarnia M, Sofronis P, Neeraj T (2011) Hydrogen interaction with multiple traps: Can it be used to mitigate embrittlement? Int J Hydrogen Energy 36:10141–10148. doi: https://doi.org/10.1016/j.ijhydene.2011.05.027

del Busto S, Betegón C, Martínez-Pañeda E (2017) A cohesive zone framework for environmentally assisted fatigue. Eng Fract Mech 185:210–226. doi: 10.1016/J.ENGFRACMECH.2017.05.021

Devanathan MA V, Stachurski Z, Beck W (1963) A technique for the evaluation of hydrogen embrittlement characteristics of electroplating baths. J Electrochem Soc 110:886–890

Díaz A, Alegre JM, Cuesta II (2016a) Coupled hydrogen diffusion simulation using a heat transfer analogy. Int J Mech Sci 115–116:. doi: 10.1016/j.ijmecsci.2016.07.020

Díaz A, Alegre JM, Cuesta II (2016b) A review on diffusion modelling in hydrogen related failures of metals. Eng Fail Anal 66:. doi: 10.1016/j.engfailanal.2016.05.019

Dietzel W, Pfuff M, Juilfs GG (2006) Hydrogen permeation in plastically deformed steel membranes. Mater Sci 42:78–84. doi: 10.1007/s11003-006-0059-8

Frappart S, Feaugas X, Creus J, et al (2012) Hydrogen solubility, diffusivity and trapping in a tempered Fe–C–Cr martensitic steel under various mechanical stress states. Mater Sci Eng A 534:384–393. doi: 10.1016/J.MSEA.2011.11.084

Frappart S, Feaugas X, Creus J, et al (2010) Study of the hydrogen diffusion and segregation into Fe–C–Mo martensitic HSLA steel using electrochemical permeation test. J Phys Chem Solids 71:1467–1479. doi: 10.1016/J.JPCS.2010.07.017

Gerberich WW, Marsh PG, Hoehn JW (1996) Hydrogen induced cracking mechanisms - are there critical experiments? In: Hydrogen Effects in Materials. pp 539–551

Gesnouin C, Hazarabedian A, Bruzzoni P, et al (2004) Effect of post-weld heat treatment on the microstructure and hydrogen permeation of 13CrNiMo steels. Corros Sci 46:1633–1647. doi: 10.1016/J.CORSCI.2003.10.006

Hirth J (1980) Effects of hydrogen on the properties of iron and steel. Metall Trans A 11:861–890. doi: 10.1007/BF02654700

Hoch BO (2015) Modelling of hydrogen diffusion in heterogeneous materials: implications of the grain boundary connectivity

Hoch BO, Metsue A, Bouhattate J, Feaugas X (2015) Effects of grain-boundary networks on the macroscopic diffusivity of hydrogen in polycrystalline materials. Comput Mater Sci 97:276–284

Jiang DE, Carter EA (2004) Diffusion of interstitial hydrogen into and through bcc Fe from first principles. Phys Rev B 70:64102

Jothi S, Croft TN, Wright L, et al (2015) Multi-phase modelling of intergranular hydrogen segregation/trapping for hydrogen embrittlement. Int J Hydrogen Energy 40:15105–15123. doi: 10.1016/J.IJHYDENE.2015.08.093

Juilfs G (2002) Das Diffusionsverhalten von Wasserstoff in einem niedriglegierten Stahl unter Berücksichtigung des Verformungsgrades. GRIN Verlag

Kharin V (2014) Comments on "Computational analysis of geometrical factors affecting experimental data extracted from hydrogen permeation tests: I – Consequences of trapping" [Int J Hydrogen Energy 36 (2011) 12644–12652] and "… II – Consequences of trapping and an oxide layer" [Int J Hydrogen Energy 37 (2012) 13574–13582], "Corrigenda …" to both [Int J Hydrogen Energy 39 (2014) 2430], and on "… III – Comparison with experimental results from the literature" [Int J Hydrogen Energy 39 (2014) 1145–1155] with "Gene. Int J Hydrogen Energy 39:19846–19850. doi: 10.1016/J.IJHYDENE.2014.09.032

Koyama M, Rohwerder M, Tasan CC, et al (2017a) Recent progress in microstructural hydrogen mapping in steels: quantification, kinetic analysis, and multi-scale characterisation. Mater Sci Technol 33:1481–1496. doi: 10.1080/02670836.2017.1299276

Koyama M, Yamasaki D, Nagashima T, et al (2017b) In situ observations of silver-decoration evolution under hydrogen permeation: Effects of grain boundary misorientation on hydrogen flux in pure iron. Scr Mater 129:48–51. doi: 10.1016/J.SCRIPTAMAT.2016.10.027

Kumnick AJ, Johnson HH (1980) Deep trapping states for hydrogen in deformed iron. Acta Metall 28:33–39. doi: http://dx.doi.org/10.1016/0001-6160(80)90038-3

Lan L, Kong X, Hu Z, et al (2016) Hydrogen permeation behavior in relation to microstructural evolution of low carbon bainitic steel weldments. Corros Sci 112:180–193. doi: 10.1016/J.CORSCI.2016.07.025

Legrand E, Feaugas X, Bouhattate J (2014) Generalized model of desorption kinetics: Characterization of hydrogen trapping in a homogeneous membrane. Int J Hydrogen Energy 39:8374–8384. doi: 10.1016/J.IJHYDENE.2014.03.191

Liu MA, Rivera-Díaz-del-Castillo PEJ, Barraza-Fierro JI, et al (2019) Microstructural influence on hydrogen permeation and trapping in steels. Mater Des 167:107605. doi: 10.1016/J.MATDES.2019.107605

Macadre A, Artamonov M, Matsuoka S, Furtado J (2011) Effects of hydrogen pressure and test frequency on fatigue crack growth properties of Ni–Cr–Mo steel candidate for a storage cylinder of a 70 MPa hydrogen filling station. Eng Fract Mech 78:3196–3211. doi: 10.1016/J.ENGFRACMECH.2011.09.007

Macadre A, Nakada N, Tsuchiyama T, Takaki S (2015) Critical grain size to limit the hydrogen-induced ductility drop in a metastable austenitic steel. Int J Hydrogen Energy 40:10697–10703. doi: 10.1016/J.IJHYDENE.2015.06.111

Martin ML, Somerday BP, Ritchie RO, et al (2012) Hydrogen-induced intergranular failure in nickel revisited. Acta Mater 60:2739–2745. doi: 10.1016/J.ACTAMAT.2012.01.040

Martínez-Pañeda E, del Busto S, Niordson CF, Betegón C (2016a) Strain gradient plasticity modeling of hydrogen diffusion to the crack tip. Int J Hydrogen Energy 41:10265–10274. doi: https://doi.org/10.1016/j.ijhydene.2016.05.014

Martínez-Pañeda E, Golahmar A, Niordson CF (2018) A phase field formulation for hydrogen assisted cracking. Comput Methods Appl Mech Eng 342:742–761. doi: 10.1016/J.CMA.2018.07.021

Martínez-Pañeda E, Niordson CF, Gangloff RP (2016b) Strain gradient plasticity-based modeling of hydrogen environment assisted cracking. Acta Mater 117:321–332. doi: 10.1016/J.ACTAMAT.2016.07.022

McMahon CJ (2001) Hydrogen-induced intergranular fracture of steels. Eng Fract Mech 68:773–788. doi: 10.1016/S0013-7944(00)00124-7

McNabb A, Foster PK (1963) A new analysis of the diffusion of hydrogen in iron and ferritic steels. Trans Metall Soc AIME 227:618–627. doi: citeulike-article-id:4956272

Miresmaeili R, Ogino M, Nakagawa T, Kanayama H (2010) A coupled elastoplastic-transient hydrogen diffusion analysis to simulate the onset of necking in tension by using the finite element method. Int J Hydrogen Energy 35:1506–1514. doi: http://dx.doi.org/10.1016/j.ijhydene.2009.11.024

Montella C (1999) Discussion on permeation transients in terms of insertion reaction mechanism and kinetics. J Electroanal Chem 465:37–50. doi: 10.1016/S0022-0728(99)00051-0

Nagao A, Smith CD, Dadfarnia M, et al (2012) The role of hydrogen in hydrogen embrittlement fracture of lath martensitic steel. Acta Mater 60:5182–5189. doi: 10.1016/J.ACTAMAT.2012.06.040

Novak P, Yuan R, Somerday BP, et al (2010) A statistical, physical-based, micro-mechanical model of hydrogen-induced intergranular fracture in steel. J Mech Phys Solids 58:206–226. doi: 10.1016/J.JMPS.2009.10.005

Olden V, Thaulow C, Johnsen R, et al (2008) Application of hydrogen influenced cohesive laws in the prediction of hydrogen induced stress cracking in 25%Cr duplex stainless steel. Eng Fract Mech

75:2333–2351. doi: https://doi.org/10.1016/j.engfracmech.2007.09.003

Ono K, Meshii M (1992) Hydrogen detrapping from grain boundaries and dislocations in high purity iron. Acta Metall Mater 40:1357–1364. doi: https://doi.org/10.1016/0956-7151(92)90436-I

Oriani RA (1970) The diffusion and trapping of hydrogen in steel. Acta Metall 18:147–157. doi: http://dx.doi.org/10.1016/0001-6160(70)90078-7

Oudriss A, Creus J, Bouhattate J, et al (2012) Grain size and grain-boundary effects on diffusion and trapping of hydrogen in pure nickel. Acta Mater 60:6814–6828. doi: https://doi.org/10.1016/j.actamat.2012.09.004

Park I-J, Lee S, Jeon H, Lee Y-K (2015) The advantage of grain refinement in the hydrogen embrittlement of Fe–18Mn–0.6C twinning-induced plasticity steel. Corros Sci 93:63–69. doi: 10.1016/J.CORSCI.2015.01.012

Raina A, Deshpande VS, Fleck NA (2017) Analysis of electro-permeation of hydrogen in metallic alloys. Philos Trans A Math Phys Eng Sci 375:20160409. doi: 10.1098/rsta.2016.0409

Sofronis P, Liang Y, Aravas N (2001) Hydrogen induced shear localization of the plastic flow in metals and alloys. Eur J Mech - A/Solids 20:857–872. doi: http://dx.doi.org/10.1016/S0997-7538(01)01179-2

Sofronis P, McMeeking RM (1989) Numerical analysis of hydrogen transport near a blunting crack tip. J Mech Phys Solids 37:317–350. doi: http://dx.doi.org/10.1016/0022-5096(89)90002-1

Song EJ (2015) Hydrogen desorption in steels. Grad Inst Ferr Technol 106

Song EJ, Suh D-W, Bhadeshia HKDH (2013) Theory for hydrogen desorption in ferritic steel. Comput Mater Sci 79:36–44. doi: 10.1016/J.COMMATSCI.2013.06.008

Takasawa K, Ikeda R, Ishikawa N, Ishigaki R (2012) Effects of grain size and dislocation density on the susceptibility to high-pressure hydrogen environment embrittlement of high-strength low-alloy steels. Int J Hydrogen Energy 37:2669–2675. doi: 10.1016/J.IJHYDENE.2011.10.099

Toribio J, Kharin V (2015) A generalised model of hydrogen diffusion in metals with multiple trap types. Philos Mag 1–23. doi: 10.1080/14786435.2015.1079660

Turnbull A (1993) Modelling of environment assisted cracking. Corros Sci 34:921–960. doi: http://dx.doi.org/10.1016/0010-938X(93)90072-O

Turnbull A (2015) Perspectives on hydrogen uptake, diffusion and trapping. Int J Hydrogen Energy. doi: http://dx.doi.org/10.1016/j.ijhydene.2015.06.147

Turnbull A, Ferriss DH, Anzai H (1996) Modelling of the hydrogen distribution at a crack tip. Mater Sci Eng A 206:1–13. doi: http://dx.doi.org/10.1016/0921-5093(95)09897-6

Turnbull A, Saenz de Santa Maria M, Thomas ND (1989) The effect of H2S concentration and pH on hydrogen permeation in AISI 410 stainless steel in 5% NaCl. Corros Sci 29:89–104. doi: 10.1016/0010-938X(89)90082-6

Van den Eeckhout E, Laureys A, Van Ingelgem Y, Verbeken K (2017) Hydrogen permeation through deformed and heat-treated Armco pure iron. Mater Sci Technol 33:1515–1523. doi: 10.1080/02670836.2017.1342015

Vecchi L, Simillion H, Montoya R, et al (2018a) Modelling of hydrogen permeation experiments in iron alloys: Characterization of the accessible parameters – Part I – The entry side. Electrochim Acta 262:57–65. doi: 10.1016/J.ELECTACTA.2017.12.172

Vecchi L, Simillion H, Montoya R, et al (2018b) Modelling of hydrogen permeation experiments in iron alloys: Characterization of the accessible parameters – Part II – The exit side. Electrochim Acta 262:153–161. doi: 10.1016/J.ELECTACTA.2017.12.173

Venezuela J, Blanch J, Zulkiply A, et al (2018) Further study of the hydrogen embrittlement of martensitic advanced high-strength steel in simulated auto service conditions. Corros Sci 135:120–135. doi: 10.1016/J.CORSCI.2018.02.037

ISO 17081:2014 Method of measurement of hydrogen permeation and determination of hydrogen uptake and transport in metals by an electrochemical technique. International Organization for Standardization, 2014, www.iso.org